%% file: sm_ichep06.tex
\documentclass[aps,prl,preprint,tightenlines,superscriptaddress,showpacs,byrevtex]{revtex4}
\usepackage{graphicx} 
\usepackage{dcolumn}  
\usepackage{verbatim}
\usepackage{amsmath}

\newcommand{\onlum}{\ensuremath{497.1 \, \mathrm{fb}^{-1}}}
\newcommand{\offlum}{\ensuremath{56.1 \, \mathrm{fb}^{-1}}}

\newcommand{\gev}{\ensuremath{\,\mathrm{GeV}}}
\newcommand{\mev}{\ensuremath{\,\mathrm{MeV}}}

\newcommand{\my}{\ensuremath{(4295\pm10^{+10}_{-3})\,\mev/c^2}}
\newcommand{\wy}{\ensuremath{(133\pm26^{+13}_{-6})\,\mev/c^2}}
\newcommand{\yy}{\ensuremath{(165\pm24^{+7}_{-23})\,\mathrm{events}}}
\newcommand{\geey}{\ensuremath{(8.7\pm1.1^{+0.3}_{-0.9})\,\mathrm{eV}}}
\newcommand{\csy}{\ensuremath{(48 \pm 6^{+1}_{-5} ) \,\mathrm{fb}}}
\newcommand{\sigy}{\ensuremath{11\sigma}}

\graphicspath{{ps}}

\begin{document}
\bibliographystyle{prsty} 

\preprint{\vbox{ \hbox{   }
                 \hbox{BELLE--CONF--0610}
}}

\title{ \quad\\[0.5cm] Study of the $Y(4260)$ resonance in $e^+e^-$ collisions with initial state
radiation at Belle }


\input{author-conf2006}

\noaffiliation

\begin{abstract}

  We present a study of $Y(4260)$ properties using the initial-state radiation
process $e^+e^- \to \gamma_{ISR} Y(4260)$. The $Y(4260)$ resonance is
reconstructed in
the $ \pi^+ \pi^-J/\psi$ decay mode, using data
collected by the
Belle detector at the KEKB $e^+e^-$ collider. We find a significant signal with a
central mass value of \my\ and a width of \wy. We find
$\Gamma_{ee}\cdot\mathcal{B}(Y(4260)\to \pi^+\pi^- J/\psi) = \geey$.
These results are preliminary.
\end{abstract}


\maketitle

\tighten

{\renewcommand{\thefootnote}{\fnsymbol{footnote}}}
\setcounter{footnote}{0}

\section{Introduction}

The Y(4260)  was originally seen by BaBar as a significant enhancement
in initial state radiation (ISR) data in
the $ \pi^+\pi^-J/\psi$ final state, where it was fitted using a
Breit-Wigner function with mass $(4259\pm8^{+2}_{-6})\,\mathrm{MeV}/c^2$ and width
$(88\pm23^{+6}_{-4})\, \mathrm{MeV}/c^2$; they found $\Gamma_{ee} \cdot
\mathcal{B}(Y(4260)\to
\pi^+\pi^-J/\psi)=(5.5\pm1.0^{+0.8}_{-0.7})\,\mathrm{eV}/c^2$~\cite{babar}.
BaBar have also seen evidence for the decay $B^-\to K^-Y(4260),\,Y(4260)\to\pi^+\pi^-J/\psi$~\cite{babar2}. The $Y(4260)$ has
since been confirmed by CLEO  in direct production $e^+e^- \to
Y(4260)$ using energy-scan data~\cite{cleo}; in addition to
observing a clear signal in $\pi^+\pi^-J/\psi$, CLEO presents
evidence for a signal in the   $ K^+K^-J/\psi$ and
$\pi^0\pi^0J/\psi$ final states. Recently, CLEO have 
also presented results using ISR data, where they find a $Y(4260) 
\to \pi^+\pi^- J/\psi$ signal~\cite{cleoisr}.
The $Y(4260)$ coincides with a minimum in the hadronic
cross-section~\cite{PDBook} and a minimum in the $D^*D^*$ cross
section~\cite{galina}; based on a fit to BES data, the authors
of Ref.~\cite{Mo:2006ss} set a lower bound
$\mathcal{B}(Y(4260)\to\pi^+\pi^-J/\psi)>0.6\%$. Such a prominent
hadronic transition to $J/\psi$ is unexpected for a
$c\overline{c}$ state above the $D{}^{(*)}\overline{D}{}^{(*)}$
threshold, and this has led to various other models being proposed. 
The $Y(4260)$ has been described as a 4-quark state~\cite{maiani-2005-72}, a
molecular
state~\cite{chiu-2006-73,yuan-2006-634,qiao-2006-639,liu-2005-72},
and a quark-gluon hybrid~\cite{zhu-2005-625,close-2005-628,kou-2005-631,luo-2006-74}; the possibility that it could be a conventional
charmonium state is discussed in
Refs~\cite{van-beveren-2006-,llanes-estrada-2005-72}. 

In this paper, we present the results of a search for
$Y(4260)\to  \pi^+\pi^-J/\psi$ production in ISR events.
This study is based on a data sample with an
integrated luminosity of \onlum at the $\Upsilon(4S)$ resonance
and \offlum collected 60 MeV/$c^2$ below the 
resonance. The  data were collected  with the Belle detector at
the KEKB asymmetric-energy $e^+e^-$ (3.5 on 8~GeV)
collider~\cite{KEKB}.

KEKB operates with a peak luminosity that exceeds
$1.6\times 10^{34}~{\rm cm}^{-2}{\rm s}^{-1}$.
The Belle detector is a large-solid-angle magnetic
spectrometer that
consists of a silicon vertex detector (SVD),
a 50-layer central drift chamber (CDC), an array of
aerogel threshold \v{C}erenkov counters (ACC),
a barrel-like arrangement of time-of-flight
scintillation counters (TOF), and an electromagnetic calorimeter
comprised of CsI(Tl) crystals (ECL) located inside
a super-conducting solenoid coil that provides a 1.5~T
magnetic field.  An iron flux-return located outside of
the coil is instrumented to detect $K_L^0$ mesons and to identify
muons (KLM).  The detector
is described in detail elsewhere~\cite{Belle}.
Two inner detector configurations were used. A 2.0 cm beampipe
and a 3-layer silicon vertex detector was used for the first sample
of 155.5 $\mathrm{fb}^{-1}$, while a 1.5 cm beampipe, a 4-layer
silicon detector and a small-cell inner drift chamber were used to record
the remaining 397.8 $\mathrm{fb}^{-1}$~\cite{Natkaniec}.

\section{Reconstruction}

\subsection{Monte Carlo and control sample}

The ISR process was simulated using Phokhara~\cite{Czyz:2005as} and
events were then ported to qq98~\cite{qq98}, and processed through a
simulation of the detector in GEANT~\cite{GEANT}.
 Decays  $Y(4260) \to \pi^+\pi^-J/\psi$ were simulated using the $\psi(2S)$ as a model~\cite{qq98}.
Events
were also generated for the  $\psi(2S)$, which is
being used as a control sample.
A data/MC efficiency correction was performed using results
from Belle particle identification studies.

\subsection{Track Selection}

Candidate events are chosen from a standard Belle data skim 
for hadronic
events: the skimming conditions are optimised for
$e^+e^-\to \Upsilon(4S) \to B\overline{B}$, not ISR events.
The skimming conditions require the presence of at least three charged tracks ($N_{\rm
ch}\ge 3$), an event vertex with radial ($r\phi$) and $z$
coordinates within 1.5 and 3.5~cm of the origin, respectively,
a total reconstructed energy in the centre-of-mass system
(CMS) greater than $0.2\sqrt{s}$
$(\sqrt{s}$ is the CMS collision energy), a $z$ component of the
net reconstructed CMS momentum less than 0.5$\sqrt{s}/c$, a total
ECL energy between 0.1$\sqrt{s}$ and 0.8$\sqrt{s}$ with at least
two energy clusters, and require $R_2$, the ratio of
second and zeroth Fox-Wolfram moments, to be less than  0.8.
(Here, the $z$ axis is aligned with the
symmetry axis of the detector, and opposite in direction to the
$e^+$ beam. The $e^-$ beam crosses this axis at an angle of
22~mrad, so the $e^+e^-$ axis as seen from the CMS is not quite
aligned with $z$: see the discussion of $\cos(\theta)$ below.)

We select charged tracks with $P_t>0.05\gev/c$ for further analysis. Muons are
identified based on track penetration depth and the hit pattern in
the KLM system; the selection has
an efficiency of $88\%$ for muons.
Electron tracks are identified using a combination of $dE/dx$ from
the CDC, ACC information, $E/p$ (E is the energy deposited in the
ECL and p is the momentum measured by the SVD and the CDC), and
track-cluster matching and shower shape in the ECL;
our selection has an efficiency of $92\%$ for electrons. To recover
radiated photons from electron candidates, the four-momenta of the
closest photon within 0.05
radians of the $e^+$ or $e^-$ direction was added to the
four-momentum of the track in subsequent analysis.
Charged pions are identified as charged tracks that fail electron and
muon selection and additionally pass a kaon veto which is applied
using energy loss measurements in the CDC, \v{C}erenkov light yields in
the ACC, and TOF information. The information from these detectors is
combined to form a $K$-$\pi$ likelihood ratio,
$\mathcal R(K/\pi) = \mathcal L_{K}/(\mathcal L_{\pi} +\mathcal
L_{K})$, where $\mathcal L_{\pi}(\mathcal L_{K})$ is the likelihood that a pion
(kaon) would produce the observed detector response.
Charged tracks with $\mathcal R(K/\pi) > 0.6$ are vetoed; this selection has
an efficiency of $95\%$ for pions, and $13\%$ for kaons.

\subsection{Event Reconstruction}

\begin{figure}[htb]
\includegraphics[width=0.54\textwidth]{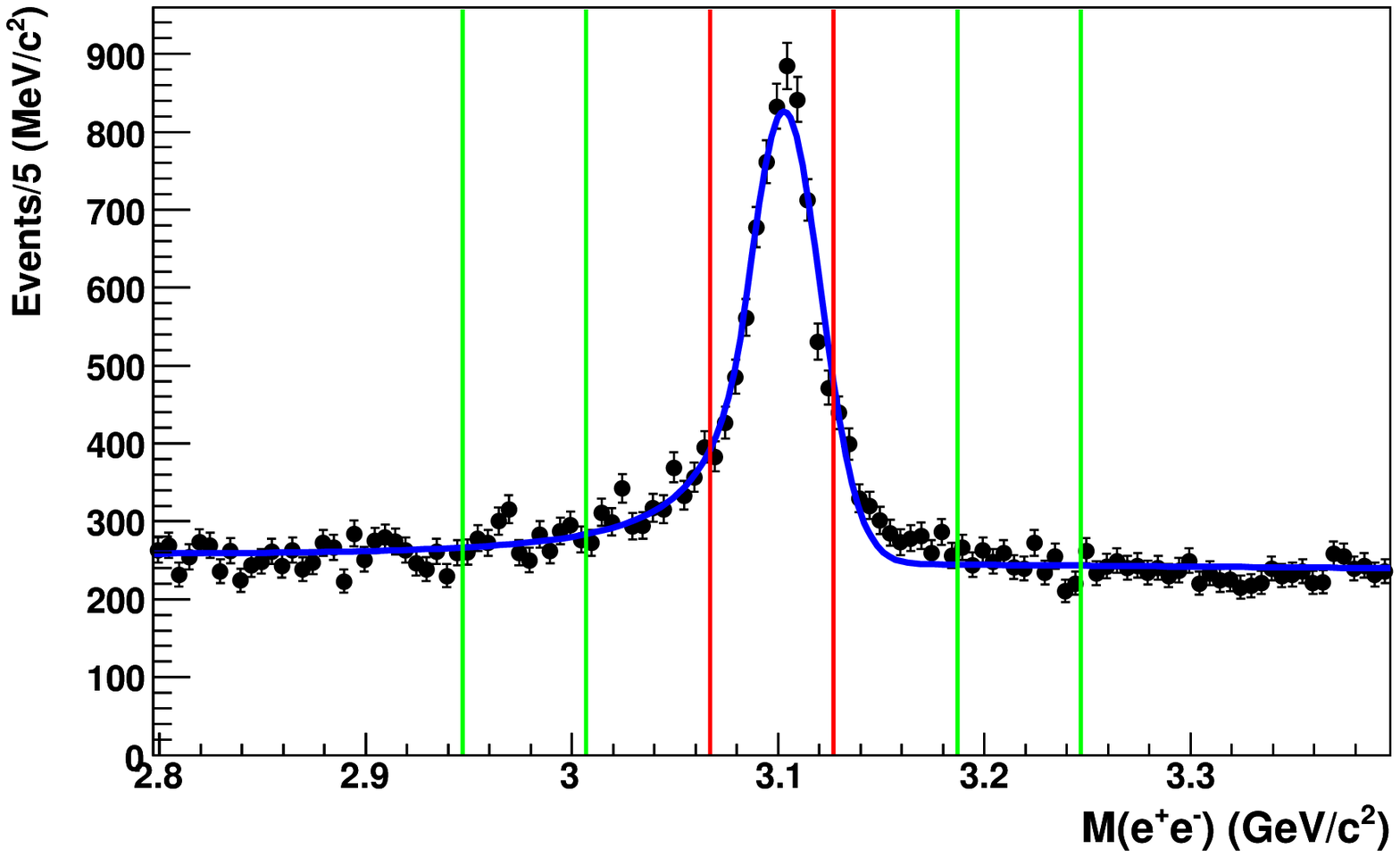}
\includegraphics[width=0.54\textwidth]{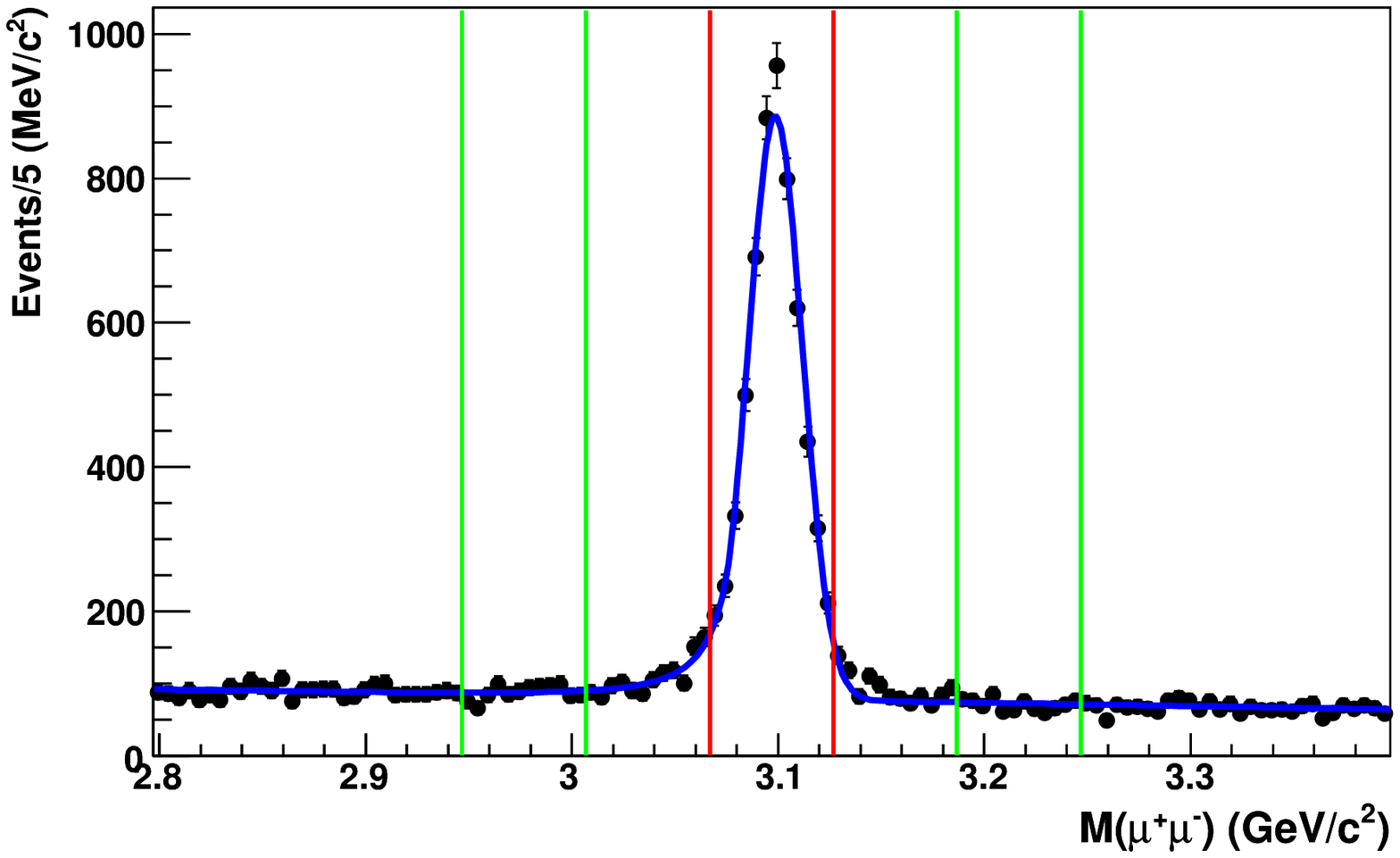}
\caption{Invariant mass spectrum of $J/\psi \to e^+e^-$ (top) and $J/\psi \to \mu^+\mu^-$ 
(bottom) candidates (points with error bars), and the fit
         described in the text (solid line). The vertical lines indicate
         the signal (red) and sideband (green) regions.}
\label{jpsiMu}
\end{figure}

We select events with four charged tracks only, then $J/\psi$
candidates are reconstructed from $J/\psi\to \ell^+\ell^-$, where $\ell$ is
either a positively identified electron or a muon; see
Fig.~\ref{jpsiMu} top and bottom, where the curves shown are fits using
the Crystal Ball function~\cite{crystalball}.
The vertical lines indicate the signal and sideband regions,
$|M(\ell^+\ell^-)-m_{J/\psi}|\in [0,30]$ and
$[90,150]\,\mathrm{MeV/}c^2$ respectively.
The $J/\psi$ candidates are then constrained to a common vertex
and the nominal $J/\psi$ mass~\cite{PDBook}, to improve the momentum
resolution. The perpendicular distance between the $J/\psi$ vertex
and the $e^+e^-$ interaction
 point is required to be less than $100\,\mu\mathrm{m}$. We then reconstruct $Y(4260)$ candidates by
combining the $J/\psi$ candidates with $\pi^+\pi^-$ pairs with
mass $M(\pi^+\pi^-)>0.4$ GeV/$c^2$. The requirement on $M(\pi^+\pi^-)$
suppresses the contribution of combinatorial background, including misidentified $\gamma \to e^+e^-$ conversions. The recoil mass squared of the
$Y(4260)$ candidates, 
\begin{equation}
M_{\rm recoil}^2 = {\left(\sqrt{s}-E_{Y}^*\right)^2-p_{Y}^{*~2}},
\end{equation}
is required to satisfy $\left|M_{\rm recoil}^2\right|<1$ $($GeV/$c^2)^2$,
in accordance with the $Y(4260)$ being produced via ISR.

\begin{figure}[htb]
\includegraphics[width=0.54\textwidth]{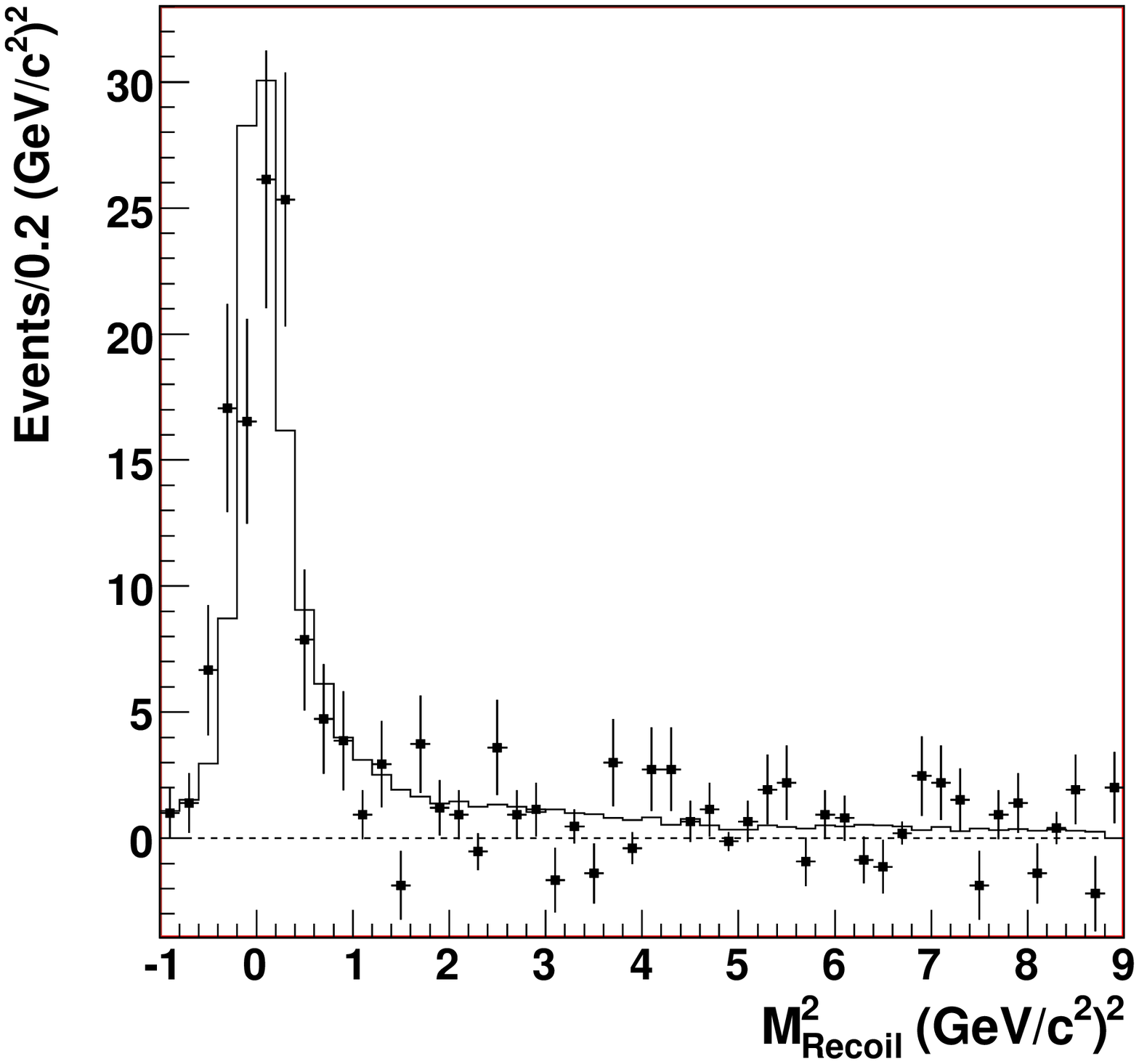}
\caption{Recoil mass squared for sideband subtracted $Y(4260)$ candidates in
        data (points) and MC (histogram).}
\label{yrec}
\end{figure}

\begin{figure}[htb]
\includegraphics[width=0.54\textwidth]{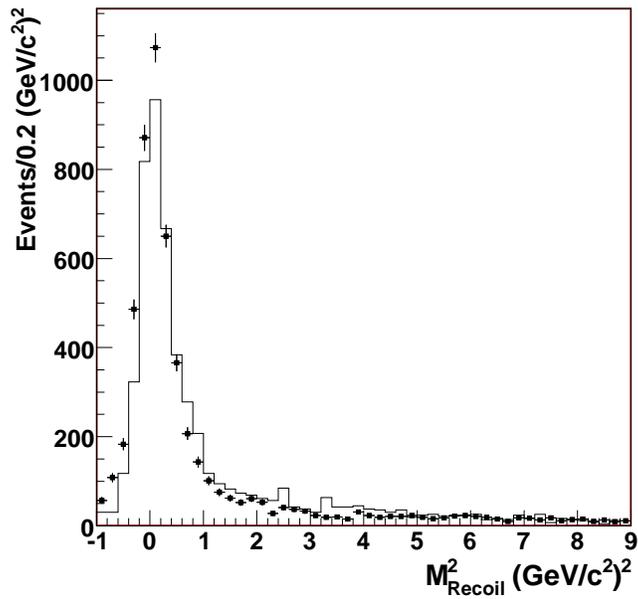}
\caption{Recoil mass squared for sideband subtracted $\psi(2S)$ candidates in
        data (points) and MC (histogram).}
\label{prec}
\end{figure}

The $\psi(2S)$ is used as a control sample and is reconstructed in the
same manner as the $Y(4260)$ candidates.
Squared recoil mass distributions are shown for the sideband subtracted
        $Y(4260)$ (Fig.~\ref{yrec})  and $\psi(2S)$ (Fig.~\ref{prec}), and
        compared with signal MC distributions, where the signal and
	sideband regions are defined in Table~\ref{sigsid}.

\begin{table}[htb]
\caption{ Signal and sideband invariant mass regions.
  $\sigma_{\psi(2S)} = 0.012$ is from the wider of the Gaussians
  fitted to the $\psi(2S)$ signal.}
\label{sigsid}
\begin{tabular}
{@{\hspace{0.5cm}}l@{\hspace{0.5cm}}||@{\hspace{0.5cm}}c@{\hspace{0.5cm}}}
\hline \hline
Region & Mass range (GeV$/c^2$) \\
\hline
 $Y(4260)$ Signal& $4.2<M(\pi^+\pi^- J/\psi)<4.4$\\
 $Y(4260)$ Sidebands & $3.9<M(\pi^+\pi^- J/\psi)<4.0$\\
 & $5.1<M(\pi^+\pi^- J/\psi)<5.5$\\
\hline 
 $\psi(2S)$ Signal& $|M(\pi^+\pi^- J/\psi)-3.686|<3\sigma=0.037$\\
 $\psi(2S)$ Sideband& $4\sigma=0.049<|M(\pi^+\pi^- J/\psi)-3.686|<0.085=7\sigma$\\
\hline \hline
\end{tabular}
\end{table}


\clearpage
\section{Fit}

We perform a binned maximum likelihood fit to the distribution of
        invariant mass $m = M(\pi^+ \pi^- J/\psi)$. The signal function is
        the product of a Breit-Wigner function, a phase space term, and
         a mass-dependent efficiency correction, added to a second-order
         polynomial representing the background. The Breit-Wigner term is
$$
\frac{1}{(m - M_Y)^2 + \frac{1}{4}\Gamma^2},
	 $$
          the phase space term is
          $$
\sqrt{((m^2+m^2_{J/\psi}-M^2_{\pi^+\pi^-})/(2m))^2-m^2_{J/\psi}},
$$
and the correction for the efficiency
         as a function of mass, based on an interpolation of results from $Y(4260)$ MC
	 samples generated with different central mass values,  is
         $$
a_{\epsilon}\cdot (m-M_0)+b_{\epsilon},
$$
where $M_0=4.3\,\mathrm{GeV}/c^2$; we fix the parameter
        $M_{\pi^+\pi^-} = 0.5\,\mathrm{GeV}/c^2$. We have neglected the effect of
interference with other resonances. The slope and intercept were
determined, respectively, to be $a_{\epsilon} = (7.4 \pm 1.3) \,(\mathrm{GeV}/c^2)^{-1}$ and
$b_{\epsilon} = (9.31 \pm 0.07)$.
The parameters allowed to float
in the fit were the number of signal ($N$) events, the mean and width of the Breit-Wigner function
    ($M_Y,\,\Gamma$), and the background parameters.

\begin{figure}[htb]
\includegraphics[width=0.54\textwidth]{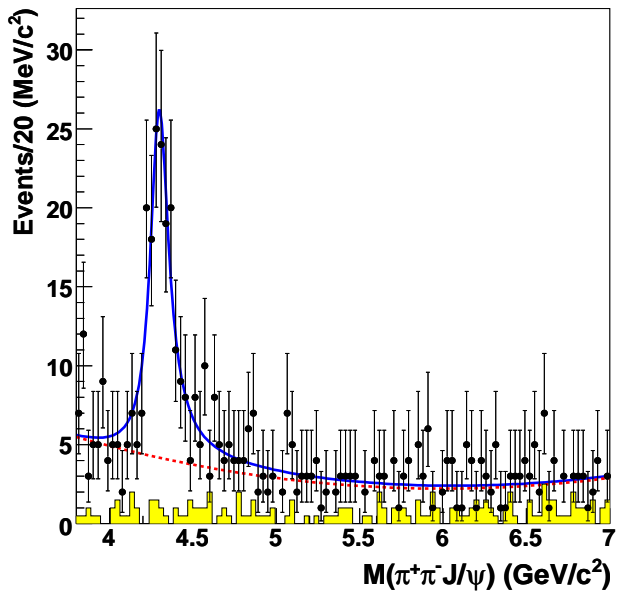}
\caption{Invariant mass for $Y(4260)$ candidates in
        data (points) and $J/\psi$ mass sidebands (shaded histogram),
	with the fit to data (solid line) and its background component (dotted
	line).}
\label{ymass}
\end{figure}

The fit to the data is shown in
Fig.~\ref{ymass}, together with the (featureless) $J/\psi$ mass
sidebands (shaded histograms). The signal parameters were found to be $M_Y=\my$, $\Gamma =\wy$ with a yield $N = \yy$, where the systematic terms
will be described in the next section. The
significance of this signal was estimated by comparing fits with
and without a signal term and was found to be \sigy.

\begin{figure}[htb]
\includegraphics[width=0.54\textwidth]{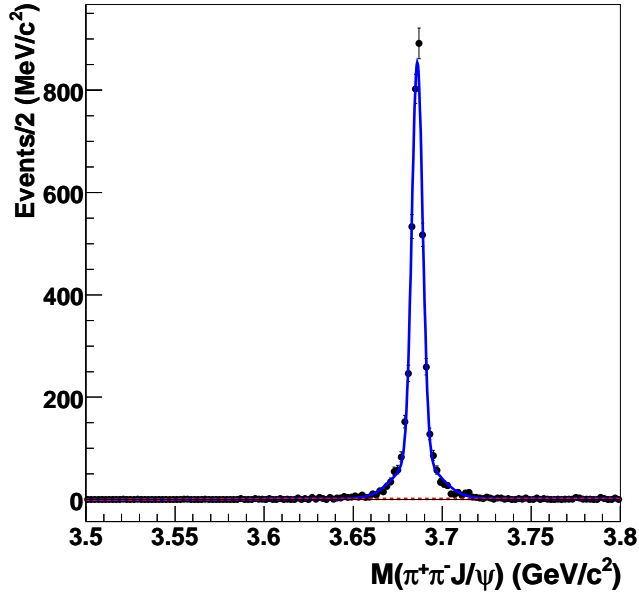}
\caption{Invariant mass for $\psi(2S)$ candidates in
        data (points), with the fit to data (solid line) and its background component (dotted
	line).}
\label{pmass}
\end{figure}

The $\psi(2S)$ control sample was fitted using two Gaussians and a
bifurcated Gaussian, with
a common mean $M_{\psi(2S)}$ floating in the fit, and a linear background: see
Fig.~\ref{pmass}.
Using the Particle Data Group value of $\Gamma_{ee}(\psi(2S))$~\cite{PDBook} we expect our $\psi(2S)$ yield to be
$4575\pm290$, where the error is dominated by MC statistics. We observe a
$\psi(2S)$ yield of $4188\pm67$, where the error is statistical only.

\subsection{ISR photon reconstruction}

\begin{figure}[htb]
\includegraphics[width=0.54\textwidth]{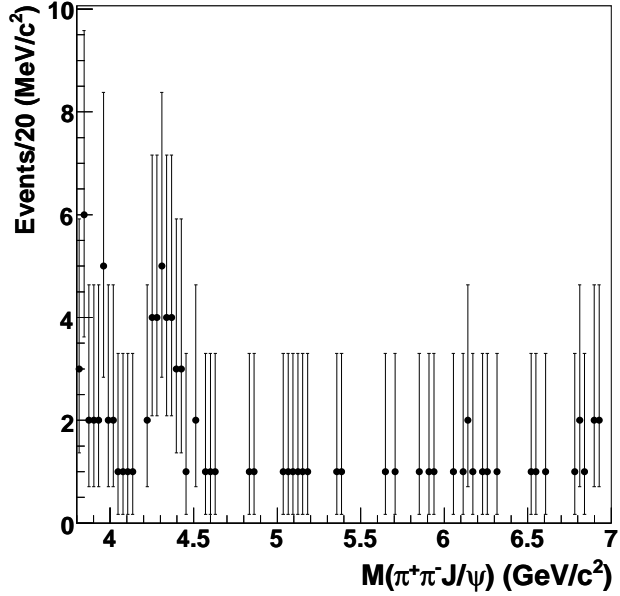}
\caption{$Y(4260)$ invariant mass spectrum requiring
reconstruction of the ISR photon.} \label{isr}
\end{figure}

Photons are identified as ECL energy clusters that are not
associated with a charged track and  have a minimum energy of
$0.060~{\rm GeV}$. If the photon with highest energy $E_{\gamma_{ISR}}$ in an event
satisfies $E_{\gamma _{ISR}}+E_{(\pi^+\pi^-J/\psi)}>10$~GeV then the ISR
photon is considered to be reconstructed. The $Y(4260)$ candidates
that pass this criteria are shown in Fig.~\ref{isr}; an enhancement
is seen at $M(\pi^+\pi^-J/\psi)\approx 4300$~MeV/$c^2$.

\section{Properties}
\subsection{Dipion Mass}
\label{q-yield-other}

\begin{figure}[htb]
  \includegraphics[width=0.54\textwidth]{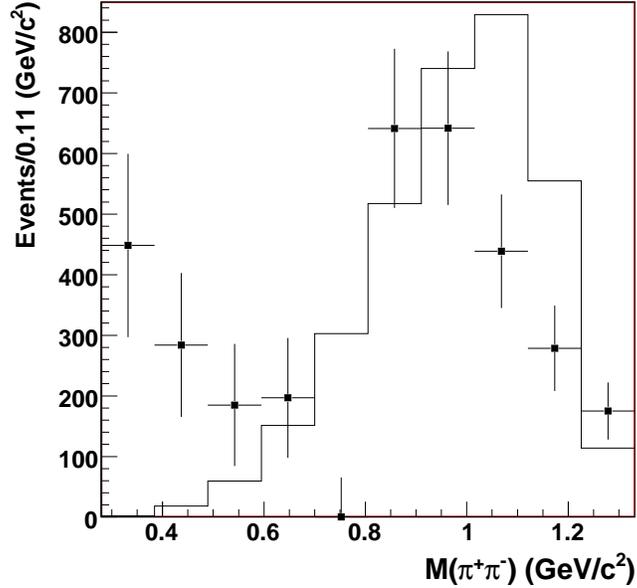}
\caption{Invariant mass of $\pi^+\pi^-$ combinations
in $Y(4260)$ candidates from fitted yields, after efficiency
correction in data (points) and MC (histogram).} \label{ydipi}
\end{figure}

\begin{figure}[htb]
  \includegraphics[width=0.54\textwidth]{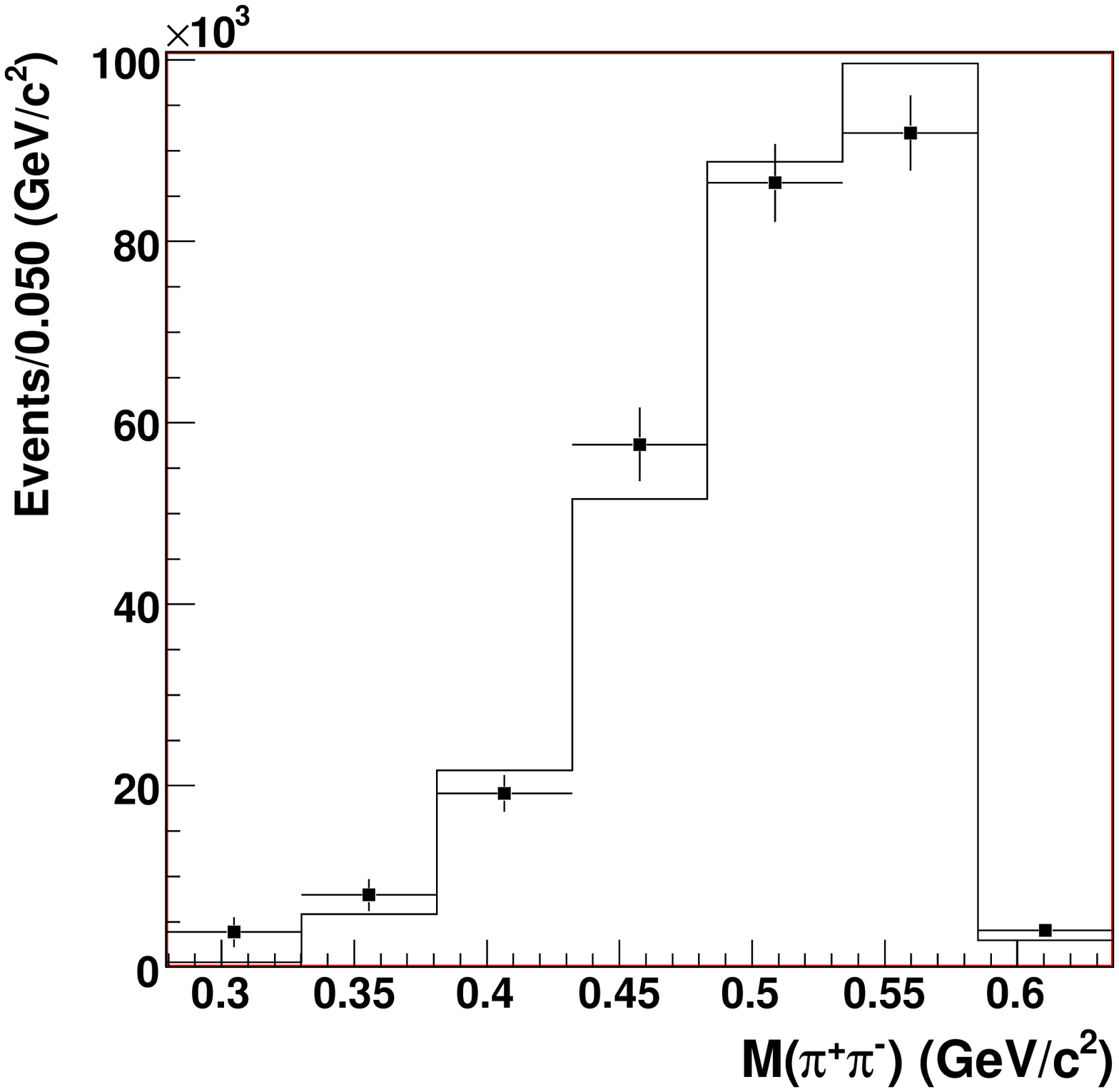}
\caption{Invariant mass of $\pi^+\pi^-$ combinations
in $\psi(2S)$ candidates, after efficiency correction in data (points) and MC (histogram).} \label{pdipi}
\end{figure}

Figure~\ref{ydipi} shows the efficiency-corrected dipion mass
    distribution for $Y(4260)$ candidates with the prediction from MC, based on fits to $M( \pi^+
    \pi^-J/\psi)$ in $M(\pi^+ \pi^-)$ bins. The
    effect of the $\pi^+\pi^-J/\psi$ mass threshold that is introduced
    by the $M(\pi^+\pi^-)$ binning is taken into account
    by multiplying
    the signal and background terms by an appropriate threshold
    function in each bin. Both signal and background parameters are
    fixed from the fit to the full sample, and only the normalisation
    and threshold function
    terms are allowed to vary. The yield ($n_i$) in each bin is corrected using the efficiency
    ($\epsilon_i$), calculated in the same bin using $Y(4260)$ MC.

The $\psi(2S)$ dipion mass distribution in Fig.~\ref{pdipi} is corrected using a 2nd order
polynomial fit to the binned efficiency from MC produced at
the nominal $\psi(2S)$ mass, the errors are dominated by
        the MC error in the binned efficiencies.

\subsection{$\cos(\theta)$ distribution}
\label{sec-ang}

\begin{figure}[htb]
  \includegraphics[height=0.44\textheight]{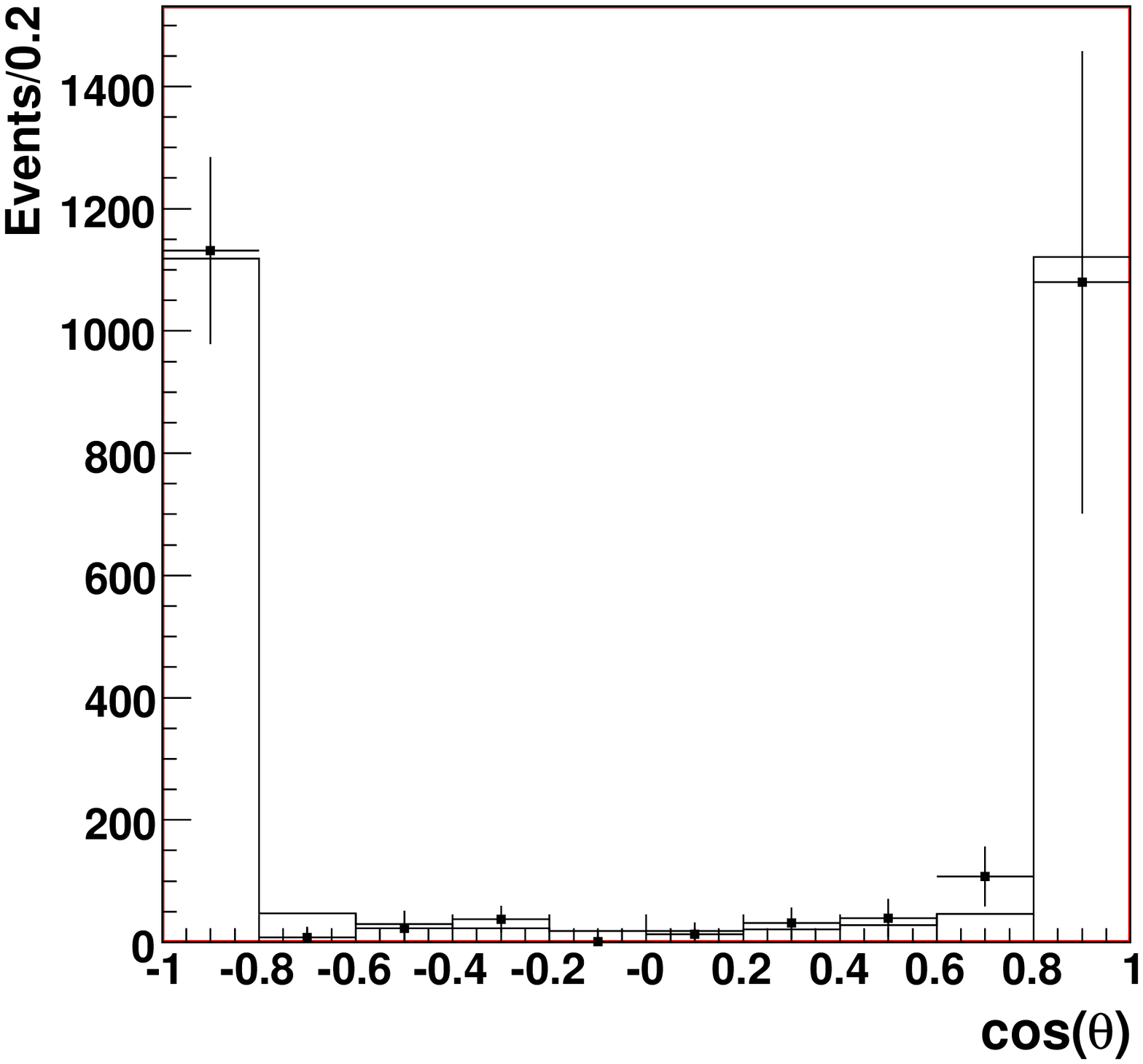}
\caption{Efficiency corrected $\cos(\theta)$ distribution for $Y(4260)$ candidates in data (points) and MC (histogram).}
\label{cospsiy0}
  \includegraphics[height=0.44\textheight]{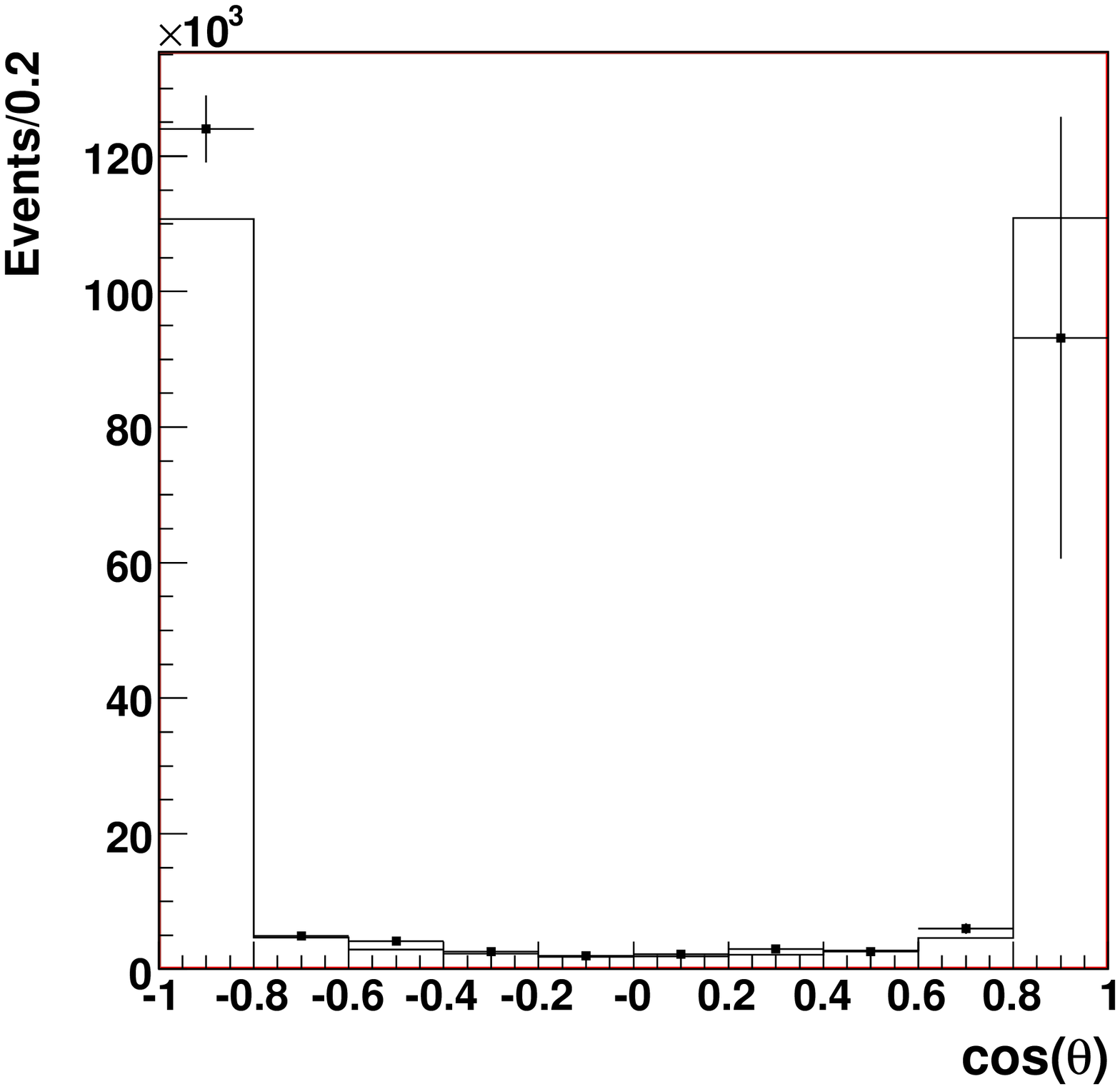}
\caption{Efficiency corrected $\cos(\theta)$ distribution for $\psi(2S)$ candidates in data (points) and MC (histogram).}
\label{cospsi}
\end{figure}

Figure~\ref{cospsiy0} shows the efficiency-corrected $\cos(\theta)$ distribution
for $Y(4260)$ candidates with the prediction from MC, where $\theta$
is the polar angle of $Y(4260)$
candidates relative to the direction of the $e^-$ beam, as seen from
    the CMS. $M( \pi^+ \pi^- J/\psi)$ fits are performed in $\cos(\theta)$ bins,
    with signal and background parameters fixed from the fit to the
    full sample, and only the normalisation terms allowed to vary.
The yield in each bin is corrected using the efficiency calculated in
the same bin, based on the $Y(4260)$ MC as above.    
     Smaller bins were used close to
$\cos(\theta) =\pm 1$, where the efficiency is changing rapidly, and
     then combined for the final plot. The same procedure was followed
    to produce the  $\cos(\theta)$ plot in the $\psi(2S)$ reference
    channel, Fig.~\ref{cospsi}.

\clearpage

\section{Systematics}
\label{sec-syst}

The systematic uncertainties on the mass and width measurement were estimated by varying the fitting
procedure. Variant fits including signal functions without
    phase space and efficiency corrections, a linear background shape,
    10 MeV/$c^2$ binning, and a (3.8,5.0) GeV/$c^2$ fit range were
    performed, and the largest deviations were taken as
    the positive and negative systematic errors due to fitting. We
    then add in quadrature terms from possible biases in $Y(4260)$
    mass reconstruction (based on a fit to MC) and the overall mass
    scale (using the $\psi(2S)$ for calibration). The fitting function does not take the variation
    with mass of the ISR cross-section into account. We estimate the
    resulting error in $\Gamma_{ee}\cdot\mathcal{B}$ (see the next section) using the
    relative change in the second-order QED radiator
    $W(s,x)$~\cite{benayoun} as the $\pi^+\pi^-J/\psi$ mass is changed
    from the fitted mean value by $\pm1\Gamma$. ($x \equiv
    2E_\gamma/\sqrt{s}$, where $E_\gamma$ is the energy of the ISR
    photon in the CMS.) 
    A summary of systematic error terms can be found in Table~\ref{sys1}.

\begin{table}[htb]
\caption{ Sources of systematic error.}
\label{sys1}
\begin{tabular}
{@{\hspace{0.5cm}}l@{\hspace{0.5cm}}||@{\hspace{0.5cm}}c@{\hspace{0.5cm}}|@{\hspace{0.5cm}}c@{\hspace{0.5cm}}|@{\hspace{0.5cm}}c@{\hspace{0.5cm}}|@{\hspace{0.5cm}}c@{\hspace{0.5cm}}}
\hline \hline
Source & Mass & Width & $\sigma\cdot\mathcal{B}$ &$\Gamma_{ee}\cdot\mathcal{B}$ \\
 &  (MeV$/c^2$)&(MeV$/c^2$) & $(\mathrm{fb})$ &$(\mathrm{eV})$ \\
\hline
Fitting procedure                     &$^{+10}_{-2}$&$^{+13}_{-6}$ &$^{+1}_{-5}$&$^{+0.2}_{-0.9}$\\
Measurement of  mass in $Y(4260)$ MC  &$\pm 2$      &--&--           &-- \\
Mass scale from measurement of $\psi(2S)$  &$\pm 2$      &--&--           &-- \\
QED radiator at masses of $\pm1\Gamma$&--           &--&--           &$\pm0.2$\\
\hline
Total                                 &  $^{+10}_{-3}$&$^{+13}_{-6}$ &$^{+1}_{-5}$&$ ^{+0.3}_{-0.9}$ \\
\hline \hline
\end{tabular}
\end{table}

Additional fits were conducted on the di-electron and di-muon
samples independently and found to be consistent with the combined
result.

\section{Cross Section}
\label{cs}

Taking the fitted $Y(4260)$ yield $n_i$ and the efficiency $\epsilon_i$
in bins of dipion mass, as described above, we identify
\begin{equation}
\begin{split}
 \sum\frac{n_i}{\epsilon_i} = \int
\mathrm{d}t \mathcal{L} \cdot 
\sigma(e^+e^- \to \gamma_{ISR} Y(4260);\; \sqrt{s}=10.58)\cdot
&\mathcal{B}(Y(4260) \to \pi^+ \pi^- J/\psi)\cdot\\
&\mathcal{B}(J/\psi \to \ell^+ \ell^-).\\
\end{split}
\label{num2}
\end{equation}
The use of $M(\pi^+\pi^-)$ bins reduces the model dependence of the
result at the cost of increasing the statistical error. 
Substituting the world average value for $\mathcal{B}(J/\psi \to \ell^+ \ell^-)$~\cite{PDBook}, we find
$$\sigma(e^+e^- \to \gamma_{ISR} Y;\;
\sqrt{s}=10.58)\cdot \mathcal{B}(Y(4260) \to \pi^+
\pi^- J/\psi) = \csy.$$
Using equation (7) from
Ref.~\cite{benayoun} we calculate, in the small width approximation, $\Gamma_{ee}\cdot
\mathcal{B}(Y(4260) \to \pi^+ \pi^- J/\psi)=\geey$.

From the signal function of a fit to the full sample (without applying an
$M(\pi^+\pi^-)$ cut) we find the
cross-section at the peak value of 4.295~GeV/$c^2$ for $e^+e^-\to Y(4260) \to
\pi^+ \pi^- J/\psi$ to be 43~pb.

\section{Discussion}

We have confirmed the $Y(4260)\to  \pi^+\pi^-J/\psi$ signal with a
significance of \sigy. The dipion mass distribution favours high values of  $M(\pi^+\pi^-)$, which is 
consistent with BaBar's result; there is also some evidence for a rise in 
cross-section near the $M(\pi^+\pi^-)$ threshold, although statistical errors 
are large.
The $\cos(\theta)$ distribution peaks strongly at values near $\pm1$, consistent
with the interpretation that the events are due to ISR, and when we require the
reconstruction of the ISR photon we see an enhancement near the
central mass value.

For our signal we find a mass of \my. There
have been predictions that a hybrid would favour decays to an
$S$-wave plus a $P$-wave state~\cite{godfrey-2006-}; the relevant mass threshold
is $M(D^0D^0_1) = 4287\,\mathrm{MeV}/c^2$. We find a mass above this
value, although the difference is not statistically significant.
CLEO's preliminary measurement is
$(4283^{+17}_{-16}\pm4)\,\mathrm{MeV}/c^2$, coinciding with the threshold
and consistent with our value; BaBar's published mass is $(4259\pm8^{+2}_{-6})\,\mathrm{MeV}/c^2$,
below threshold and $2.6\sigma$ below our value.
Our measurement of the width, $\Gamma=\wy$ is consistent
with those of CLEO $(70^{+40}_{-25}\pm5)$~MeV$/c^2$ and BaBar
$(88\pm23^{+6}_{-4})\, \mathrm{MeV}/c^2$; errors are large.
Our result for $\Gamma_{ee}\cdot{\cal B}(Y\rightarrow\pi^+\pi^- J/\psi)$
is about $1.6\sigma$ higher than BaBar's reported value $(5.5\pm1.0^{+0.8}_{-0.7})\,\mathrm{eV}/c^2$.

\section{Summary}

We have investigated the properties of the $Y(4260)$ using the initial-state radiation
process $e^+e^- \to \gamma_{ISR} Y(4260)$. We observe a significant
signal described by a Breit-Wigner with mass \my\ and width  \wy, and  find
$\Gamma_{ee}\cdot\mathcal{B}(Y(4260)\to \pi^+\pi^- J/\psi) = \geey$.
These results are preliminary.

\section{Acknowledgements}
We thank the KEKB group for the excellent operation of the
accelerator, the KEK cryogenics group for the efficient operation
of the solenoid, and the KEK computer group and the National
Institute of Informatics for valuable computing and Super-SINET
network support. We acknowledge support from the Ministry of
Education, Culture, Sports, Science, and Technology of Japan and
the Japan Society for the Promotion of Science; the Australian
Research Council and the Australian Department of Education,
Science and Training; the National Science Foundation of China and
the Knowledge Innovation Program of the Chinese Academy of
Sciences under contract No.~10575109 and IHEP-U-503; the
Department of Science and Technology of India; the BK21 program of
the Ministry of Education of Korea, the CHEP SRC program and Basic
Research program (grant No.~R01-2005-000-10089-0) of the Korea
Science and Engineering Foundation, and the Pure Basic Research
Group program of the Korea Research Foundation; the Polish State
Committee for Scientific Research;
the Ministry of Science and Technology of the Russian
Federation; the Slovenian Research Agency;  the Swiss
National Science Foundation; the National Science Council
and the Ministry of Education of Taiwan; and the U.S.\
Department of Energy.

\clearpage

\bibliography{ichep-bib} 

\end{document}

%% file: author-conf2006.tex
\affiliation{Budker Institute of Nuclear Physics, Novosibirsk}
\affiliation{Chiba University, Chiba}
\affiliation{Chonnam National University, Kwangju}
\affiliation{University of Cincinnati, Cincinnati, Ohio 45221}
\affiliation{University of Frankfurt, Frankfurt}
\affiliation{The Graduate University for Advanced Studies, Hayama} 
\affiliation{Gyeongsang National University, Chinju}
\affiliation{University of Hawaii, Honolulu, Hawaii 96822}
\affiliation{High Energy Accelerator Research Organization (KEK), Tsukuba}
\affiliation{Hiroshima Institute of Technology, Hiroshima}
\affiliation{University of Illinois at Urbana-Champaign, Urbana, Illinois 61801}
\affiliation{Institute of High Energy Physics, Chinese Academy of Sciences, Beijing}
\affiliation{Institute of High Energy Physics, Vienna}
\affiliation{Institute of High Energy Physics, Protvino}
\affiliation{Institute for Theoretical and Experimental Physics, Moscow}
\affiliation{J. Stefan Institute, Ljubljana}
\affiliation{Kanagawa University, Yokohama}
\affiliation{Korea University, Seoul}
\affiliation{Kyoto University, Kyoto}
\affiliation{Kyungpook National University, Taegu}
\affiliation{Swiss Federal Institute of Technology of Lausanne, EPFL, Lausanne}
\affiliation{University of Ljubljana, Ljubljana}
\affiliation{University of Maribor, Maribor}
\affiliation{University of Melbourne, Victoria}
\affiliation{Nagoya University, Nagoya}
\affiliation{Nara Women's University, Nara}
\affiliation{National Central University, Chung-li}
\affiliation{National United University, Miao Li}
\affiliation{Department of Physics, National Taiwan University, Taipei}
\affiliation{H. Niewodniczanski Institute of Nuclear Physics, Krakow}
\affiliation{Nippon Dental University, Niigata}
\affiliation{Niigata University, Niigata}
\affiliation{University of Nova Gorica, Nova Gorica}
\affiliation{Osaka City University, Osaka}
\affiliation{Osaka University, Osaka}
\affiliation{Panjab University, Chandigarh}
\affiliation{Peking University, Beijing}
\affiliation{University of Pittsburgh, Pittsburgh, Pennsylvania 15260}
\affiliation{Princeton University, Princeton, New Jersey 08544}
\affiliation{RIKEN BNL Research Center, Upton, New York 11973}
\affiliation{Saga University, Saga}
\affiliation{University of Science and Technology of China, Hefei}
\affiliation{Seoul National University, Seoul}
\affiliation{Shinshu University, Nagano}
\affiliation{Sungkyunkwan University, Suwon}
\affiliation{University of Sydney, Sydney NSW}
\affiliation{Tata Institute of Fundamental Research, Bombay}
\affiliation{Toho University, Funabashi}
\affiliation{Tohoku Gakuin University, Tagajo}
\affiliation{Tohoku University, Sendai}
\affiliation{Department of Physics, University of Tokyo, Tokyo}
\affiliation{Tokyo Institute of Technology, Tokyo}
\affiliation{Tokyo Metropolitan University, Tokyo}
\affiliation{Tokyo University of Agriculture and Technology, Tokyo}
\affiliation{Toyama National College of Maritime Technology, Toyama}
\affiliation{University of Tsukuba, Tsukuba}
\affiliation{Virginia Polytechnic Institute and State University, Blacksburg, Virginia 24061}
\affiliation{Yonsei University, Seoul}
  \author{K.~Abe}\affiliation{High Energy Accelerator Research Organization (KEK), Tsukuba} 
  \author{K.~Abe}\affiliation{Tohoku Gakuin University, Tagajo} 
  \author{I.~Adachi}\affiliation{High Energy Accelerator Research Organization (KEK), Tsukuba} 
  \author{H.~Aihara}\affiliation{Department of Physics, University of Tokyo, Tokyo} 
  \author{D.~Anipko}\affiliation{Budker Institute of Nuclear Physics, Novosibirsk} 
  \author{K.~Aoki}\affiliation{Nagoya University, Nagoya} 
  \author{T.~Arakawa}\affiliation{Niigata University, Niigata} 
  \author{K.~Arinstein}\affiliation{Budker Institute of Nuclear Physics, Novosibirsk} 
  \author{Y.~Asano}\affiliation{University of Tsukuba, Tsukuba} 
  \author{T.~Aso}\affiliation{Toyama National College of Maritime Technology, Toyama} 
% ...
\collaboration{The Belle Collaboration}

%% file: sm_ichep06.bbl
\begin{thebibliography}{10}

\bibitem{babar}
B.~Aubert \emph{et al.} (BaBar collaboration), Physical Review Letters
  \textbf{95}, 142001 (2005).

\bibitem{babar2}
B. Aubert {\it et~al.}, Physical Review D {\bf 73},  011101  (2006).

\bibitem{cleo}
T.E.~Coan \emph{et al.} (CLEO collaboration), Physical Review Letters
  \textbf{96}, 162003 (2006).

\bibitem{cleoisr}
Q.~He {\it et al.} (CLEO Collaboration), hep-ex/0611021 (2006); submitted to
  Physical Review D as a Rapid Communication.

\bibitem{PDBook}
W.-M.~Yao {\it et al.} (Particle Data Group), Journal of Physics G {\bf 33}, 1
  (2006).

\bibitem{galina}
K.~Abe {\it et al.} (Belle Collaboration), hep-ex/0608018 (2006).

\bibitem{Mo:2006ss}
X.~H. Mo {\it et~al.}, Physics Letters B {\bf 640},  182  (2006).

\bibitem{maiani-2005-72}
L. Maiani, F. Piccinini, A.~D. Polosa, and V. Riquer, Physical Review D {\bf
  72},  031502  (2005).

\bibitem{chiu-2006-73}
T.~C. T.-W. Chiu and T.-H. Hsieh, Physical Review D {\bf 73},  094510  (2006).

\bibitem{yuan-2006-634}
C.~Z. Yuan, P. Wang, and X.~H. Mo, Physics Letters B {\bf 634},  399  (2006).

\bibitem{qiao-2006-639}
C.-F. Qiao, Physics Letters B {\bf 639},  263  (2006).

\bibitem{liu-2005-72}
X. Liu, X.-Q. Zeng, and X.-Q. Li, Physical Review D {\bf 72},  054023  (2005).

\bibitem{zhu-2005-625}
S.-L. Zhu, Physics Letters B {\bf 625},  212  (2005).

\bibitem{close-2005-628}
F.~E. Close and P.~R. Page, Physics Letters B {\bf 628},  215  (2005).

\bibitem{kou-2005-631}
E. Kou and O. Pene, Physics Letters B {\bf 631},  164  (2005).

\bibitem{luo-2006-74}
X.-Q. Luo and Y. Liu, Physical Review D {\bf 74},  034502  (2006).

\bibitem{van-beveren-2006-}
E. {van Beveren} and G. Rupp, hep-ph/0605317, 2006.

\bibitem{llanes-estrada-2005-72}
F.~J. Llanes-Estrada, Physical Review D {\bf 72},  031503  (2005).

\bibitem{KEKB}
S.~Kurokawa and E.~Kikutani, Nuclear Instruments and Methods A {\bf 499}, 1
  (2003), and other papers included in this volume.

\bibitem{Belle}
A.~Abashian {\it et al.} (Belle Collaboration), Nuclear Instruments and Methods
  A {\bf 479}, 117 (2002).

\bibitem{Natkaniec}
Z.~Natkaniec \emph{et al.} (Belle SVD2 Group), Nuclear Instruments and Methods
  A {\bf 560} 1 (2006).

\bibitem{Czyz:2005as}
H. Czyz {\it et~al.}, European Physics Journal C {\bf 47},  617  (2006).

\bibitem{qq98}
http://www.lns.cornell.edu/public/CLEO/soft/QQ/.

\bibitem{GEANT}
R.~Brun \emph{et al.}, GEANT 3.21, CERN Report DD/EE/84--1 (1984).

\bibitem{crystalball}
T.~Skwarnicki, Ph.D.~Thesis, Institute for Nuclear Physics, Krakow 1986; DESY
  Internal Report, DESY F31-86-02 (1986).

\bibitem{benayoun}
M.~Benayoun \emph{et al.}, Modern Physics Letters A \textbf{14}, 2605 (1999).

\bibitem{godfrey-2006-}
S. Godfrey, hep-ph/0605152, 2006.

\end{thebibliography}
